\documentclass[useAMS,usenatbib]{mn2e}
\usepackage{graphicx}
\usepackage{color}
\usepackage{soul}

\title[Deep spectroscopic LF of Abell\,85]{Deep spectroscopic luminosity function of Abell\,85: no evidence for a steep upturn of the  faint-end slope}
\author[I. Agulli et al.]{I. Agulli$^{1,2,4,5}$, J. A. L. Aguerri$^{1,2}$,  R. S\'anchez-Janssen$^{3}$, R. Barrena$^{1,2}$,
\newauthor
 A. Diaferio$^{4,5}$, A. L. Serra$^{4,5,6}$, J. M\' endez-Abreu$^{7}$\\
$^1$ Instituto de Astrofisica de Canarias, C/ Via Lactea s/n, 38200 La Laguna, Tenerife, Spain, \\
$^2$ Departamento de Astrofisica, Universidad de La Laguna, E-38205 La Laguna, Tenerife, Spain,\\
$^3$NRC Herzberg Institute of Astrophysics, 5071 West Saanich Road, Victoria, V9E2E7, Canada\\
$^4$ Dipartimento di Fisica, Universit\`a di Torino, Via P. Giuria 1, I-10125, Torino, Italy\\
$^5$ Istituto Nazionale di Fisica Nucleare (INFN), sezione di Torino, Via P. Giuria 1, I-10125, Torino, Italy\\
$^6$ Istituto Nazionale di Astrofisica (INAF), Osservatorio Astronomico di Torino, Strada Osservatorio 20, I-10025 Pino Torinese, Torino, Italy\\
$^7$ School of Physics \& Astronomy, University of St Andrews, St Andrews KY16 9AJ, Fife, Scotland, UK}

\begin{document}

\date{Accepted . Received 2014 April 30; in original form 2013 November}

\pagerange{\pageref{firstpage}--\pageref{lastpage}} \pubyear{2002}

\maketitle

\label{firstpage}

\begin{abstract}
We present a new deep determination of the spectroscopic LF within the virial radius of the nearby and massive Abell\,85 (A85) cluster down to the dwarf regime (M* + 6) using VLT/VIMOS spectra for $\sim 2000$ galaxies with m$_r \leq 21$ mag and $\langle \mu_{e,r} \rangle \leq 24$ mag arcsec$^{-2}$. The resulting LF from 438 cluster members is best modelled by a double Schechter function due to the presence of a statistically significant upturn at the faint-end. The amplitude of this upturn ($\alpha_{f} = -1.58^{+0.19}_{-0.15}$), however, is much smaller than that of the SDSS composite photometric cluster LF by \cite{popesso2006},  $\alpha_{f} \sim$ -2. The faint-end slope of the LF in A85 is consistent, within the uncertainties, with that of the field. The red galaxy population dominates the LF at low luminosities, and is the main responsible for the upturn. The fact that the slopes of the spectroscopic LFs in the field and in a cluster as massive as A85 are similar suggests that the cluster environment does not play a major role in determining the abundance of low-mass galaxies.
\end{abstract}


\section{Introduction}
The  galaxy luminosity function (hereafter, LF)  represents  the  number density  of galaxies of a  given luminosity. It is a robust observable, extensively used in the past, to study the properties of galaxy populations \citep[e.g.][and references therein]{blanton2003}. The comparison between the behaviour of the LF and the halo mass function at faint magnitudes has been proposed as a crucial test to understand galaxy formation processes. Indeed, the cold dark matter theory predicts halo mass functions with a slope of $\sim-1.9$ \citep[e.g.][and reference therein]{springel2008}, steeper than the one observed in deep LFs of nearby galaxy clusters, or in the field \citep[$\sim -1.1:-1.5$][]{trenthamt2002,depropris2003,blanton2005}. This is the so-called missing satellite problem, well reported in censuses of galaxies around the Milky Way and M31 \citep[e.g.][]{klypin1999}. Attempts to solve these discrepancies invoke several physical mechanisms that halt the star formation and darken or destroy dwarf galaxies, including high gas cooling times \citep{wr1978}, and suppression of low-mass galaxies due to a combination of feedback, photoionization or/and dynamical processes \citep{benson2002,benson2003,brooks2013}.
 
Claims of the environmental dependence of the LF are numerous \citep[e.g.][]{tully2002,trentham2002,trentham2005,infante2003}, but the exact shape and significance of this dependence is still a matter of debate. Many previous studies suggest that the most striking differences between low-density and high-density environments concern the faint-end slope of the LF, with cluster galaxies showing a higher abundance of low-luminosity galaxies than the field \citep[e.g.][and references therein]{blanton2005, popesso2006}. This has important implications, as it suggests that, whatever the mechanisms preventing the formation of galaxies within low-mass satellites are, they depend on the host halo mass. \cite{tully2002} suggest that perhaps reionization inhibited the collapse of gas in late-forming dwarfs in low-density environments, whereas a larger fraction of low-mass haloes formed before the photoinization of the intergalactic medium in overdensities that ultimately become clusters. Interestingly, this mechanism could be at the heart of the discrepancies existing in the literature about the faint-end of the LF in clusters. While some authors observe a marked upturn at faint magnitudes \citep[][]{yagi2002,popesso2006,barkhouse2007}, others find a more regular behaviour \citep[e.g.][]{sanchez2005,andreon2006}. When observed, this upturn is usually due to early-type dwarfs located in the outer regions of clusters, suggesting that cluster-related mechanisms may be responsible for the paucity of dwarfs observed in the denser inner regions \citep{pfeffer2013}. Moreover, the slopes derived in the upturn region are rather steep, as much as to be fully consistent with that of the halo mass function. Unfortunately, there is no evidence for the existence of such a dramatic steeping in the few cluster LFs where spectroscopic or visual membership has been determined \citep[][]{hilker2003,rines2003,mieske2007,misgeld2008,misgeld2009}. More extensive spectroscopic cluster surveys are needed to derive robust results on the faint-end slope of the LF, and in this way put constraints on the role played by the environment on the formation of baryonic structures within low-mass dark matter halos.

We have undertaken a project in order to obtain spectroscopy of galaxies in nearby clusters down to the dwarf regime ($M > M^*+6$). This dataset will allow us to infer accurate cluster membership and analyze several properties of the dwarf galaxy population in nearby clusters, minimizing the contribution of background sources. In the present work, we present a study of the spectroscopic LF of  (A85), a nearby ($z = 0.055$) and massive cluster \citep[$M_{200} = 2.5 \times 10^{14} \; M_{\odot}$, and $R_{200} = 1.02 \, h^{-1} \,$Mpc][]{rines2006}. This cluster is not completely virialized, since several substructures and infalling groups have been identified \citep{durret1999,aguerri2010,cava2010}. A85 is an ideal target for a deep study of the LF, as spectroscopy within the virial radius is almost complete down to m$_r \sim$ 18, resulting in 273 confirmed members \citep[][]{aguerri2007}. The new dataset presented here reaches three mag fainter, and almost doubles the number cluster members. 

Throughout this work we have used the cosmological parameters $H_0 = 75 \; \mathrm{km} \, \mathrm{s}^{-1} \mathrm{Mpc}^{-1}$, $\Omega _m = 0.3$ and $\Omega _{\Lambda} = 0.7$. 

\section{The observational data on A85}
\subsection{Deep VLT/VIMOS Spectroscopy}
Our parent photometric catalogue contains all galaxies brighter than $m_r = 22$ mag\footnote{The apparent magnitudes used are the dered SDSS-DR6 $r$-band magnitudes.} from the SDSS-DR6 \citep[][]{adelman2008}, and within $R_{200}$\footnote{The cluster center is assumed to be at the brightest cluster galaxy (BCG, $\alpha$(J2000): $00^h \, 41^m \, 50.448^s$ $\delta$(J2000): $-9^{\circ} \, 18'  \, 11.45''$). This is a sensible assumption because the peak of X-ray emission lies at only 7 kpc from the BCG \citep{popesso2004}.}. Figure \ref{cmd} shows the colour-magnitude diagram of A85 for the galaxies included in this catalogue. The target galaxies for our spectroscopic observations were selected among those with no existing redshift in the literature and bluer that $g-r = 1.0$ (see Fig. \ref{cmd}). This is the colour of a 12 Gyr old stellar population with [Fe/H] = +0.25 supersolar metallicity \citep{worthey1994}, typical of very luminous early-type galaxies. As a result, this colour selection should minimize the contamination by background sources, while matching at the same time the colour distribution of galaxies in the nearby Universe \citep[e.g.][]{hogg2004}. 

The observations were carried out using the multi-object-spectroscopy (MOS) mode of VLT/VIMOS, in combination with the LR-blue+OS-blue grisms and filters (Program 083.A-0962(B), PI R. S\'anchez-Janssen). To maximize the number of targets, and avoid the gaps between the instrument CCDs, we designed 25 masks with large overlaps covering an area of 3.0 $\times$ 2.6 Mpc$^2$ around the central galaxy in A85 -- i.e. extending out to more than 1\,R$_{200}$. This observational strategy allowed us to obtain 2861 low-resolution spectra (R=180) of galaxies down to $m_r= 22$ mag. We exposed during 1000 s to reach a signal-to-noise ($S/N$) in the range $6 - 10$ down to the limiting magnitude. The data were reduced using GASGANO and the provided pipeline \citep{izzo2004}. The spectra were wavelength calibrated using the HeNe lamp, which yield a wavelength accuracy  of $\sim 0.5$ $\rm \AA$ pixel$^{-1}$ in the full spectral range ($3700 - 6700 \, \rm \AA$).

\subsection{Redshift Determination and Cluster Membership}
The recessional velocity of the observed galaxies were determined using the \textit{rvsao.xcsao} \textit{IRAF} task \citep{kurtz1992}. This task cross correlates a template spectrum \citep[in this work][]{kennicutt1992} with the observed galaxy spectrum. This technique allowed us to determine the recessional velocity in 2070 spectra. The remaining spectra had too low $S/N$ to estimate reliable redshifts. \textit{xcsao} formal errors are smaller than true intrinsic errors \citep[e.g.][]{bardelli1994}, and reliable errors can only be estimated by observing galaxies more than once. Our observational strategy allowed us to obtain 676 repeated spectra result in a one-sigma velocity uncertainty of $\sim 500$ km s$^{-1}$. The redshifts from the literature (SDSS-DR6 and NED catalogues), together with our new data, result in a total number of 1593 galaxy velocities in the A85 direction within R$_{200}$ and $14  < m_r < 22$. 

\begin{figure}
\centering
  \includegraphics[width=1\linewidth]{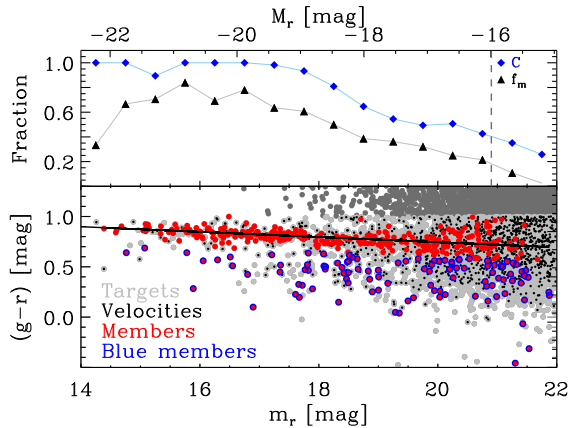}
  \caption{Lower panel: colour-magnitude diagram of the galaxies in the direction of A85. Grey points are the target galaxies. Red and blue symbols show red and blue cluster members, respectively. The solid line represents the red sequence of the cluster. Upper panel: spectroscopic completeness  ($C$, diamonds) and cluster member function ($f_m$, black triangles) as a function of $r$-band magnitude. The dashed vertical line represents our limit magnitude for the spectroscopic LF.}
\label{cmd}
\end{figure}

The caustic method \citep{diaferio1997,diaferio1999,serra2011} estimates the escape velocity and the mass profile of galaxy clusters in both their virial and infall regions, where the assumption of dynamical equilibrium does not necessarily hold. A by-product of the caustic technique is the identification of the members of the cluster with an interloper contamination of only 2\% within R$_{200}$, on average \citep{serra2013}. The application of the caustic technique to our spectroscopic catalogue resulted in a sample of  $434$ cluster members within $R_{200}$, 284 of which are new data. 

We define the completeness of our data as $C = N_z / N_{phot}$, with $N_z$ being the number of measured redshift and $N_{phot}$ the number of photometric targets. Figure \ref{cmd} shows that $C$ is higher than 90 $\%$ for galaxies with M$_r < -19$ and it decreases around $40 \, \%$ at $M_r \sim -16$. We also defined the member fraction as $f_m = N_m / N_z$, being $N_m$ the number of members. The member fraction also strongly depends on luminosity. Thus, $f_m$ is higher than 80 $\%$ down to M$_r < -19$ and then rapidly decreases down to $\sim$ 20 $\%$ at   M$_r = -16$ (see Fig. \ref{cmd}).

\section{The  spectroscopic  LF  of  A85}  
The A85 LF is computed using all cluster members with m$_r \leq 21$ mag and $\langle \mu_{e,r} \rangle \leq 24$ mag arcsec$^{-2}$. These limits correspond to the values where the galaxy counts stop increasing, and thus determine our completeness limits. We note that our uniform spectroscopic selection function (cf. Sect. 2.1) does not introduce any bias in magnitude, $\langle \mu_{e,r} \rangle $, or colour.

\begin{figure}
\includegraphics[width=1\linewidth]{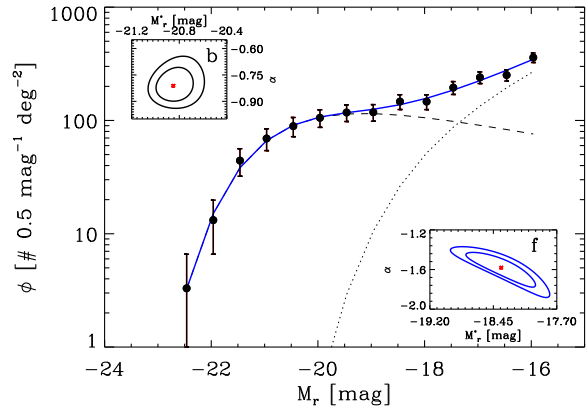}
\caption{Black points are the observed spectroscopic LF of A85. The solid blue line shows the fit of the LF by a double Schechter. The bright ($b$) and faint ($f$) components of the fit are represented by a dashed and dotted lines, respectively. The 68$\%$ and 99$\%$ c.l. for the fitted parameters are shown in the insets.}
\label{LF}
\end{figure}

Figure \ref{LF} shows the $r$-band spectroscopic LF of A85. It is computed as $\phi(M_r) = N_{phot}(M_r) \times  f_m(M_r) / (0.5 \times A)$, where $A$ is the observed area and 0.5 is the magnitude bin. 
The Pearson test shows that the observed LF cannot be modelled by a single Schechter at 99$\%$ confidence level, due to the presence of a statistically significant upturn at $M_r > -18$ (see Fig. \ref{LF}). The observed LF is better parameterized using two Schechter functions. Because of the degeneracy in fitting a double Schechter profile, we followed a two-steps process \citep[see][]{barkhouse2007}. First, we fitted the bright part ($b: -22.5<$ $M_{r}<-19.0$) of the LF obtaining $M^*_b$ and $\alpha_b$. Then, these parameters were fixed when a double Schechter fit was performed to the total LF. Table \ref{tabella_data} shows the M$_r^*$ and $\alpha$ values of the faint ($f: -19.0<$ $M_{r}<-16.0$) and bright parts of the best-fit LF. 

\begin{figure}
\includegraphics[width=1\linewidth]{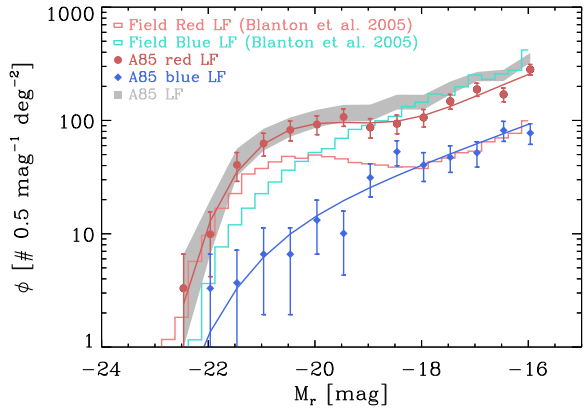}
\caption{The spectroscopic LF of A85 (black points), blue and red diamonds show the LF of blue and red galaxies of A85. The full lines correspond to the Schecter function fits. The histograms are the LF of field red and blue galaxies (Blanton et al. 2005).}
\label{LF_rb}
\end{figure}

In order to better understand the nature of the galaxies responsible for the upturn at the faint end of the LF, we classified galaxies in blue and red ones according to their $(g-r)$ colours. Thus, blue galaxies are those with $(g-r) < (g-r)_{RS} - 3 \sigma $ and red ones are the remaining cluster members; where $(g-r)_{RS}$ is the colour of the red sequence of A85, and $\sigma_{RS}$ represents its dispersion\footnote{The red sequence of the cluster and its dispersion were measured in the magnitude range $-22.5 < $ M$_r < -19.0$.} (see Fig. \ref{cmd}). Figure \ref{LF_rb} shows the spectroscopic LF of the blue and red populations of A85. Naturally, red galaxies completely dominate in number. The red LF departs from the single Schechter shape, showing the characteristic flattening at intermediate luminosities followed by a (mild) upturn at the faint end. The blue LF, however, is well fitted by a single Schechter function. This is in qualitative agreement with previous work \citep[e.g.][P06 hereafter]{popesso2006}. Nevertheless, it is remarkable that the faint-end slopes of both the red and blue populations are virtually identical (see Table \ref{tabella_data}).

\begin{table}
\begin{center}
\caption{Schechter Function Parameters.\label{tabella_data}}
\begin{tabular}{cccc}
\hline\hline
 mag interval &$M_{r}^* $ [mag] &$\alpha $\\
\hline
-22.5 $<$ M$_{r}$ $<$ -19.0  &$-20.85\; ^{+0.14}_{-0.14}$  &$-0.79\; ^{+0.08}_{-0.09}$ \\
-19.0 $<$ M$_{r}$ $<$ -16.0   &$-18.36\; ^{+0.41}_{-0.40}$  &$-1.58\; ^{+0.19}_{-0.15}$ \\
\hline
&     red &\\
\hline
-22.5 $<$ M$_{r}$ $<$ -19.0  &$-20.71\; ^{+0.13}_{-0.15}$  &$-0.63\; ^{+0.09}_{-0.08}$ \\
-19.0 $<$ M$_{r}$ $<$ -16.0    &$-17.90\; ^{+0.35}_{-0.19}$  &$-1.46\; ^{+0.18}_{-0.17}$ \\
\hline
&      blue &\\
\hline
-22.5 $<$ M$_{r}$ $<$ -16.0   &$-21.29 \; ^{+0.36}_{-0.44}$  &$-1.43 \; ^{+0.06}_{-0.05}$ \\
\hline
\end{tabular}
\end{center}
\end{table}

\section{Discussion}
In Fig.3 we additionally show the LFs of blue and red field populations from \cite{blanton2005} \footnote{Throughout this work, all the literature LFs are normalized so that they have the same number of counts as our A85 LF in the magnitude range $-22 < M_{r} < -19$.}. Their field LF  is best described by a double Schechter function with faint-end slope -1.5. 

\begin{figure}
\includegraphics[width=1\linewidth]{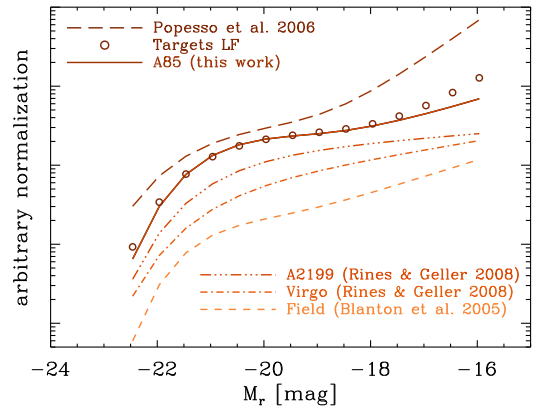}
\caption{Comparison between our LF and others present in literature: the stacked photometric LF of 69 clusters from Popesso et al. (2006), the spectroscopic LFs of A2199 and Virgo from Rines \& Geller (2008), and the field galaxy LF from Blanton et al. (2005). The photometric LF of A85 is also shown with open circles.}
\label{LF_cfr}
\end{figure}

Figure \ref{LF_cfr} shows the comparison between our LF and other cluster and field LFs from the literature. We compare with the LF of the Virgo and Abell\,2199 clusters \citep{rines2008} because they are the two deepest spectroscopic LFs of nearby clusters covering a significant fraction of their respective virial radii (out to $\sim 0.7\,R_{200}$, very similar to our coverage). In addition, we include the LF from P06 because it is a photometric LF obtained by stacking a large sample of 69 clusters, and it exhibits the most statistically meaningful example of an upturn. Finally, we compare with the field LF from \cite{blanton2005}. We note that all LFs in Fig.4 have been derived using SDSS photometry. The very steep upturn present by the stacked photometric LF  ($\alpha \sim -2.2$) is not observed in any of the nearby clusters. While A85 shows a slightly steeper slope than Virgo and A2199, it is not nearly as steep as the photometric LF. 

The discrepancy between the LF of A85 and the composite cluster LF from P06 can be traced back to the different methodologies applied to derive them. First, we introduced a stringent colour cut in the selection of our spectroscopic targets (see Sect.2.1). Second, our spectroscopic LF is based on accurate cluster membership using the galaxy recessional velocity, while photometric LFs necessarily rely on a statistical background subtraction. The open circles in Fig. \ref{LF_cfr} show the photometric LF of A85, derived from the number counts of candidate cluster galaxies with $(g-r) < 1.0$, and after performing a statistical background subtraction using galaxies (with the same colour cut) from 50 SDSS random fields of the same area as our A85 survey. It is clear that the photometric LF of A85 closely matches the spectroscopic LF, except for a steeper faint-end slope, which moves towards the P06 value. We quantify this difference by fitting a power law function to these LFs in the $ -18 \leq M_{r} \leq -16$ magnitude range. We prefer this simple approach to minimize the degeneracies present when fitting a double Schechter function (see Sect.3). We derive power-law faint-end slopes $\alpha_{f} = -1.5, -1.8,$ and  $-2.1$ for the spectroscopic LF of A85, the photometric LF of A85, and the stacked photometric LF from  P06, respectively. We note that cosmic variance of the cluster LF, or any mass dependence, can not explain the discrepancy between the composite cluster LF and that of A85. P06 show that 90 per cent of their nearby cluster sample have individual LFs consistent  with the composite one shown in Fig.\ref{LF_cfr}. Moreover, the photometric LF for A85 itself from P06 is totally consistent with having a very steep faint end (see their Figure 5). This all suggests that the composite photometric LF from  P06 suffers from contamination at the faint end, resulting in much steeper slopes than what is found using spectroscopic data.

The field spectroscopic LF presented by \citep[][]{blanton2005} shows an upturn and a faint-end slope ($\alpha=-1.5$) similar to the one of A85. Contrary to some claims from photometrically derived cluster LFs, Fig. \ref{LF_cfr} shows that the faint-end slope in A85 and the field are consistent with each other -- i.e. clusters of these masses do not seem to contain a significant excess of dwarf galaxies with respect to the field. This, in turn, suggests that the environment may not play a major role in determining the abundance of low-mass galaxies \citep[cf.][]{tully2002}, but only acts to modify their star formation activity. While the overall abundance of dwarfs in the field and in clusters like A85 is the same, blue dwarfs dominate in the former environments, and red dwarfs in the latter.

Environmental processes involving the loss of the galaxy gas reservoirs \citep[e.g.][]{quilis2000, bekki2002}, followed by the subsequent halt of the star formation, and the reddening of their stellar populations are the obvious mechanisms invoked to explain the colour transformation of cluster dwarfs.
It is however not yet clear whether the efficiency of these processes depends on the halo mass or not. \cite{lisker2013} propose that quiescent cluster dwarfs originate as the result of early \citep[see also][] {sanchez2012} \emph{and} prolonged environmental influence in \emph{both} group- and cluster-size haloes. Along these lines, \cite{wetzel2013} find that group preprocessing is responsible for up to half of the quenched satellite population in massive clusters. On the other hand, P06 suggest that the excess of red dwarfs in clusters is a threshold process that occurs within the cluster virial radius \citep[see also][]{sanchez2008}, but their exact abundance is nevertheless a function of clustercentric radius: the upturn becomes steeper as the distance from the centre increases. 
Our dataset calls for a study of the LF as a function of radial distance from the cluster centre, and an investigation of the properties of A85 dwarfs in substructures and infalling groups. This will be the subject of future  papers in this series. 

\section{Conclusions}
We obtained 2861 low resolution spectra for galaxies down to $m_{r}=22$ mag within the virial radius of the nearby ($z=0.055$) galaxy cluster A85. This unique dataset allowed us to identify 438 galaxy cluster members, and build the spectroscopic LF of A85 down to M* + 6. The resulting LF is best modelled by a double Schechter function due to the presence of a statistically significant upturn at the faint-end. The amplitude of this upturn ($\alpha = -1.58^{+0.19}_{-0.15}$), however, is much smaller than that of most photometric cluster LFs previously reported in the literature. The faint-end slope of the LF in A85 is consistent, within the uncertainties, with that of the field. We investigate the nature of the galaxy population dominating the faint end of the LF of A85 by  dividing the galaxies according to their $(g-r)$ colour. The red population dominates at low luminosities, and is the main responsible for the upturn. This is different from the field LF: while a relatively steep upturn is also present in the red population, blue galaxies are the prevalent population. The fact that the slopes of the spectroscopic LFs in the field and in a cluster as massive as A85 are similar suggests that the cluster environment does not play a major role in determining the abundance of low-mass galaxies, but it does influence the star formation history of dwarfs. 

\textbf{\textit{Acknowledgements.}} 
IA, AD and ALS acknowledge partial support from the
INFN grant InDark and from the grant Progetti di Ateneo TO Call 2012 0011 ‘Marco Polo’
of the University of Torino.

\end{document}